\documentclass{article}
\usepackage{spconf,amsmath,graphicx}
\usepackage{xcolor}
\usepackage{hyperref}
\usepackage{hyperref}
\usepackage{hyperref}
\usepackage[normalem]{ulem}
\usepackage{booktabs}

\usepackage{tikz}
\usetikzlibrary{arrows}
\title{Augmented Contrastive Self-Supervised Learning for Audio Invariant Representations}
\name{ Melikasadat Emami$^\dagger$\thanks{Work done while Melikasadat Emami was an intern at Microsoft.},  Dung Tran$^\diamondsuit$,  Kazuhito Koishida$^\diamondsuit$}
\address{$^\dagger$ ECE Department, University of California,  Los Angeles, Los Angeles, CA\\ $^\diamondsuit$ Applied Sciences Group, Microsoft, Redmond, WA }
\newcommand{\xt}{\widetilde{x}}

\newcommand{\tran}{^{\text{\sf T}}}

\newcommand{\Ac}{\mathcal{A}}

\begin{document}
\maketitle
\begin{abstract}
Improving generalization is a major challenge in audio classification due to labeled data scarcity. Self-supervised learning (SSL) methods tackle this by leveraging unlabeled data to learn useful features for downstream classification tasks. In this work, we propose an augmented contrastive SSL framework to learn invariant representations from unlabeled data. Our method applies various perturbations to the unlabeled input data and utilizes contrastive learning to learn representations robust to such perturbations. Experimental results on the Audioset and DESED datasets show that our framework significantly outperforms state-of-the-art SSL and supervised learning methods on sound/event classification tasks.

 \end{abstract}
\section{Introduction}
\label{sec:intro}
Improving generalization is a key challenge in supervised learning for audio classification \cite{zhang2021understanding}. Traditional approach to solve this problem is to obtain more labeled training data to learn invariant features via shared labels. However, data labeling is a highly expensive process and audio recordings are usually weakly labeled. This leads to a limited availability of desired labeled data to train models in a supervised fashion.

Recently, self-supervised learning (SSL) methods remedy the labeled data scarcity issue by leveraging unlabeled data to learn useful representations and transfer the pretrained features in subsequent supervised tasks \cite{bachman2019learning, oord2018representation, chen2020simple}. In this approach, \emph{invariance} of the pretrained features to transformations is crucial to downstream tasks. Recent works have proposed methods to encourage invariance for vision tasks. However, for audio applications, learning invariant representations without using labeled data remains a challenge.

Inspired by recent advances in SSL for audio, we propose an augmented contrastive learning framework that learns invariant features using unlabeled data. In this approach, the embeddings are learnt by comparing multiple augmented audio segments from the same or different recordings in a contrastive fashion: The model encourages the similarity between the representations learnt from the same audio recording and discourages the similarity between those from different audio recordings. We incorporate various types of perturbations to the input data and minimize a contrastive loss to learn representations robust to such pertubations. Utilizing a vast amount of unlabeled data, this produces representations invariant to nuisance factors and improve generalization of the supervised downstream task given the same amount of labeled data. Our experimental results show that our finetuned model outperformed the supervised network and a naive SSL approach for the sound-event classification on the Audioset \cite{45857} and DESED \cite{Turpault2019} datasets.

\textbf{Prior work.} 
Recently proposed self-supervised learning frameworks in vision and audio have shown great promise in learning useful representations for downstream tasks.
Most of these frameworks use contrastive learning approaches that rely on negative sampling \cite{bachman2019learning, oord2018representation, chen2020simple, he2020momentum, caron2020unsupervised, saeed2021contrastive}. 
Some, however, avoid the negative samples by introducing architectural changes to compensate for that \cite{NEURIPS2020_f3ada80d, niizumi2021byol}. In vision, besides altering the network architecture, different augmentation techniques have been used to encourage feature invariance, and delivered better performance than fully supervised training \cite{chen2020simple, he2020momentum,caron2020unsupervised}. Inspired by this success, several SSL methods \cite{saeed2021contrastive, jansen2018unsupervised, fonseca2021unsupervised, ravanelli2020multi} have been presented to exploit massive unlabeled data to learn audio representations.

The most related framework to this work is COLA \cite{saeed2021contrastive} which learns general-purpose audio representations in a contrastive fashion. However, COLA has no mechanism to enforce feature invariance. This explains why, despite using a vast amount of unlabeled data for pre-trained features, its results are only comparable or slightly better than supervised learning given the same labeled dataset as shown in our experimental results.

\textbf{Outline.} The rest of the paper is organized as follows. In section \ref{sec:model}, we introduce our augmented contrastive learning framework and its modules. There, we describe the augmentation strategies used in our work. In section \ref{sec:exp}, we present the experimental results and show that our augmented framework outperforms state-of-the-art SSL and supervised training.

\section{Contrastive Learning Framework for Audio/Speeech} \label{sec:model}
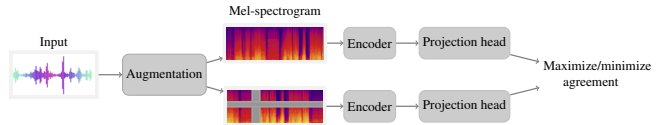
\begin{figure}
        \centering
        \hspace*{-2cm}    
        \scalebox{.28}{%

\begin{tikzpicture}[thick,->,draw=black!80, node distance=2.5cm, scale=1, every node/.style={scale=1.5, font=\large}]
\tikzstyle{every pin edge}=[<-,line width=3pt]
\tikzstyle{neuron}=[circle,fill=black!25,minimum size=15pt,inner sep=0pt]

\tikzstyle{noise}=[circle,draw=blue!25, fill=blue!20,minimum size=15pt,inner sep=0pt]

\tikzstyle{annot} = [text width=1em, text centered]
\tikzstyle{null} = [draw = none , fill= none]
\tikzstyle{line} = [draw, black, -latex']

\tikzstyle{sound block}=[rectangle,draw=black!5, fill=black!5,minimum size=40pt, rounded corners=0mm, minimum height=.5 cm,minimum width=0.5cm]

\tikzstyle{Aug block}=[rectangle,draw=black!25, fill=black!20,minimum size=40pt, rounded corners=3mm, minimum height=1cm,minimum width=1.5cm]

\tikzstyle{Aug2 block}=[rectangle,draw=black!35, fill=black!20,minimum size=40pt, rounded corners=3mm, minimum height=1.3cm,minimum width=1.5cm]

\tikzstyle{spec block}=[rectangle,draw=black!5, fill=black!5,minimum size=40pt, rounded corners=1mm, minimum height=1.2cm,minimum width=1cm]

\node[sound block, xshift=0cm, yshift= 0cm ](input){\includegraphics[scale=1]{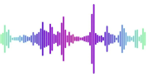}};
\node[annot, above of = input, yshift = -1.5cm, text width = 10cm]{Input};

\node[Aug2 block, right of = input , xshift = 1cm, yshift = 0cm](aug){Augmentation};
 
\path (input) edge [above, line width=0.6mm, draw=black!50] (aug);

\node[spec block, right of = aug , xshift = 1cm, yshift = 1cm](anchor_spec){\includegraphics[scale=1]{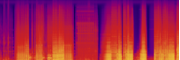}};

\node[spec block, right of = aug , xshift = 1cm, yshift = -1cm](masked_spec){\includegraphics[scale=1]{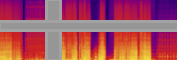}};

\node[Aug block, right of = anchor_spec , xshift = 0.5cm, yshift = 0cm](enc1){Encoder};

\node[Aug block, right of = masked_spec , xshift = 0.5cm, yshift = 0cm](enc2){Encoder};

\path (aug) edge [above, line width=0.6mm, draw=black!50] (anchor_spec);
\path (aug) edge [above, line width=0.6mm, draw=black!50] (masked_spec);
\path (anchor_spec) edge [above, line width=0.6mm, draw=black!50] (enc1);
\node[annot, above of =anchor_spec, xshift= 0cm,  yshift = -1.5cm, text width = 10cm]{Mel-spectrogram};

\path (masked_spec) edge [above, line width=0.6mm, draw=black!50] (enc2);

\node[Aug block, right of = enc1 , xshift = 0.5cm, yshift = 0cm](prj1){Projection head};

\node[Aug block, right of = enc2 , xshift = 0.5cm, yshift = 0cm](prj2){Projection head};

\path (enc1) edge [above, line width=0.6mm, draw=black!50] (prj1);
\path (enc2) edge [above, line width=0.6mm, draw=black!50] (prj2);

\node[annot, right of = aug, xshift= 11cm,  yshift = 0cm, text width = 3cm](max){ Maximize/minimize agreement};

\path (prj1) edge [above, line width=0.6mm, draw=black!50] (max);
\path (prj2) edge [above, line width=0.6mm, draw=black!50] (max);

\end{tikzpicture}
}
        \caption{Contrastive learning framework for audio/speech signals}
        \label{fig:framework}
\end{figure}

The general contrastive learning pipeline is illustrated in Figure \ref{fig:framework}. 
We learn audio/speech representations by pretraining a network using unlabeled data. As shown in Figure \ref{fig:framework}, the pipeline includes: 1) an augmentation module, 2) an Encoder network with a projection head, and 3) a loss function. 

During pretraining of the contrastive network, we maximize the similarity between latent representations of the augmented segments from the same audio clip (positive pair) and minimize that between augmented representations of different clips (negative pairs). The augmentation module serves as a regularizer encouraging the contrastive network to learn invariant representations. The contrastive feature extractor is later used in a downstream task such as sound event classification.

\textbf{Augmentation:} 
This module provides a series of transformations to the input audio/speech signal. Given a batch of size $B$ and for any signal from the batch, the module produces $B$ pairs of data including one positive pair and $B-1$ negative pairs. Let $\{a_i,a'_i\}_{i=1}^{B} \in \Ac$, where $\Ac$ is the set of transformations applied by this module, and $\{x_i\}_{i=1}^B$ the selected batch. We denote 
\begin{equation*}
    \xt_i = a_i(x_i), \quad \xt'_i = a'_i(x_i) \quad i = 1,\dots,B. 
\end{equation*}
The positive and negative pairs are then define as follows:
\begin{align*}
    (\xt_i , \xt'_i)&: \text{ positive pairs} \quad i = 1,\dots,B\\
    (\xt_i , \xt'_j)&: \text{ negative pairs} \quad j = 1,\dots,B, j\neq i. 
\end{align*}
The augmentation module outputs a positive or negative pair of transformed signals, a pair of mel-spectrograms in our case, that is fed to the encoder. We discuss our proposed transformations in detail in section \ref{subsec:augmentation}.

\textbf{Encoder:}
In our model, the feature extractor network is a convolutional encoder which receives mel-spectrograms as its input. The extracted features are passed through a projection network, a multi-layer perceptron (MLP), and then used in the contrastive loss. Let $f(.)$ be the encoder network that maps each input into the latent space. Further assume that $g(.)$ represents the projection head. For each pair $(\xt,\xt')$ of inputs, we notate 
\begin{equation}
            h = f(\xt),\  h' = f(\xt'), \quad \quad 
            z = g(h),\   z' = g(h').
\end{equation} 
We follow the COLA's \cite{saeed2021contrastive} framework for the feature extractor network architecture in our work. The extracted features are then used for optimizing the contrastive loss function via a similarity measure.

\textbf{Similarity measure:}
As mentioned earlier, we aim to maximize the similarity between two augmented segments of the same sound clip and minimize the similarity for different ones. Common approaches for calculating the similarity measure include dot product and cosine similarity. In our framework, we choose bilinear transformation. This type of similarity has been used before in \cite{saeed2021contrastive, oord2018representation}. The score is calculated as 
\begin{equation}
     s(x,x') = z\tran W z',
\end{equation}
where $W$ are the bilinear parameters.

\textbf{Loss function:}
Our contrastive loss function is the multi-class cross entropy over the similarity scores. For a given $x_i$, the loss is equivalent to
\begin{equation}
    \mathcal{L} = - \log \frac{e^{s(\xt_i,\xt_i')}}{e^{s(\xt_i,\xt_i')} + \sum_{j\neq i}e^{s(\xt_i,\xt_j')}}.
\end{equation}
This loss is closely related to the Info-NCE loss.

\subsection{Proposed Augmentation Module}\label{subsec:augmentation}
\begin{figure}
        \centering
         \hspace*{-1.5cm} 
        \scalebox{.26}{%

\begin{tikzpicture}[thick,->,draw=black!80, node distance=2.5cm, scale=1, every node/.style={scale=1.5, font=\large}]
\tikzstyle{every pin edge}=[<-,line width=3pt]
\tikzstyle{neuron}=[circle,fill=black!25,minimum size=15pt,inner sep=0pt]

\tikzstyle{noise}=[circle,draw=blue!25, fill=blue!20,minimum size=15pt,inner sep=0pt]

\tikzstyle{annot} = [text width=1em, text centered]
\tikzstyle{null} = [draw = none , fill= none]
\tikzstyle{line} = [draw, black, -latex']

\node[rectangle, fill = black!5, draw=black!10, minimum width = 23cm , minimum height= 5.5cm, rounded corners=5mm, xshift = 0cm, yshift = 0cm](augmentation){};

\node[annot, above of = augmentation, yshift = -0.2cm, text width = 10cm]{Augmentation};

\tikzstyle{sound block}=[rectangle,draw=black!5, fill=black!5,minimum size=40pt, rounded corners=0mm, minimum height=.5 cm,minimum width=0.5cm]

\tikzstyle{Aug block}=[rectangle,draw=black!25, fill=black!20,minimum size=40pt, rounded corners=3mm, minimum height=1cm,minimum width=1.5cm]

\tikzstyle{spec block}=[rectangle,draw=black!5, fill=black!5,minimum size=40pt, rounded corners=1mm, minimum height=1.2cm,minimum width=1cm]

\node[sound block, right of = augmentation , xshift=-12.5cm, yshift= 0cm ](input){\includegraphics[scale=1]{input.png}};
 \node[annot, above of = input, yshift = -1.5cm, text width = 10cm]{Input};
\node[sound block, right of = input , xshift = 1.2cm, yshift = 0.8cm](anchor){\includegraphics[scale=1]{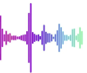}};
\node[Aug block, right of = input , xshift = 1.3cm, yshift = -1cm](speed){$\begin{aligned} & \text{ Time Stretching} \\ &(\times 1.1) \text{or} (\times 0.9)\end{aligned}$};

\node[sound block, right of = speed , xshift = 0.4cm, yshift = 0cm](pos){\includegraphics[scale=1]{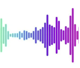}};

\path (input) edge [above, line width=0.6mm, draw=black!50] (anchor);
\path (input) edge [above, line width=0.6mm, draw=black!50] (speed);

\node[Aug block, right of = anchor , xshift = 0.3cm, yshift = 0cm](noise){Noise};

\path (anchor) edge [above, line width=0.6mm, draw=black!50] (noise);
\path (speed) edge [above, line width=0.6mm, draw=black!50] (pos);

\node[Aug block, right of = pos , xshift = 0.3cm, yshift = 0cm](rir){RIR Filtering};

\path (pos) edge [above, line width=0.6mm, draw=black!50] (rir);
\node[spec block, right of = rir , xshift = 0.6cm, yshift = 0cm](spec){\includegraphics[scale=1]{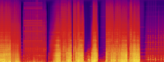}};

\node[Aug block, right of = spec , xshift = 0.8cm, yshift = 0cm](mask){$\begin{aligned} & \text{ Time/Freq  }\\ & \text{ Masking } \end{aligned}$};

\path (rir) edge [above, line width=0.6mm, draw=black!50] (spec);
\path (spec) edge [above, line width=0.6mm, draw=black!50] (mask);
\node[spec block, right of = mask , xshift = 1.1cm, yshift = 0cm](masked_spec){\includegraphics[scale=1]{masked_spec.png}};
\path (mask) edge [above, line width=0.6mm, draw=black!50] (masked_spec);
\node[spec block, right of = noise , xshift = 10.5cm, yshift = 0cm](anchor_spec){\includegraphics[scale=1]{anchor_spec.png}};
\path (noise) edge [above, line width=0.6mm, draw=black!50] (anchor_spec);

\end{tikzpicture}
}
         \caption{Proposed augmentation module}
         \label{fig:augmentation}
\end{figure}

Unlike traditional augmentation techniques that aim to enlarge the training dataset using artificial samples, we incorporate augmentation and contrastive training to learn representations invariant to nuisance factors.
In particular, our proposed augmentation module generates positive and negative pairs for the contrastive network. This module contains a series of transformations that, via the contrastive loss, preserve useful information and ignore nuisance factors in the input audio/speech signals. The input to the module is a 10-second audio/speech clip. We randomly choose and add noise to a 1-second chunk of this audio. We then calculate the mel-spectrogram of this audio signal and construct our anchor feature. The positive feature is another 1-second segment, from the same clip, that is passed through the augmentation blocks. Finally, we choose a 1-second from any other audio clip, which is different from the clip that we sample the anchor and positive feature, in the batch as the negative feature. This negative sample is also fed to the augmentation blocks. Figure \ref{fig:augmentation} illustrates different transformations, described below, in the augmentation module.

\textbf{Time stretching:} This block randomly (with probability $=0.5$) speeds up or slows down the original audio clip in time by a factor of $0.1$. It makes the network robust to temporal variations in the input. Note that for the positive and negative pairs, we choose the 1-sec segments after this block.

\textbf{RIR filtering:} This module passes the 1-sec segment through a room impulse response filter chosen from \cite{reddy2021interspeech}. The transformation leads to contrastive features invariant to different recording environments.

\textbf{Time/Freq masking:} This operation applies a time and frequency masking to the calculated mel-spectrograms. It encourages the contrastive network to learn representations robust to large distortions in the input.

Even though each augmentation type leads to discriminative features benefiting certain classes in the downstream classification task, our ablation study indicates that a combination of some of these transformations produces the highest classification performance on average for all classes in the considered dataset.

\subsection{Downstream Tasks}
We perform downstream tasks for sound/event classification on both DESED \cite{Turpault2019} and Audioset \cite{45857} datasets. We aim to improve the generalization error using the proposed augmentation module. As both of the test datasets are unbalanced, meaning the labels are represented unequally, standard evaluation metrics such as class accuracy are unreliable and misleading. Therefore, we use the F1-score and weighted average precision (wAP) score to evaluate the model performance. These metrics are calculated based on the precision and recall scores defined below:
\begin{align}
    \text{precision} &= \frac{\text{TP}}{\text{TP}+\text{FP}}, \quad \text{recall} = \frac{\text{TP}}{\text{TP}+\text{FN}}, \nonumber \\
    \text{F1-score} &= \frac{2\times \text{precision} \times \text{recall}}{\text{precision}+\text{recall}}.
\end{align}
Here, TP, FP, and FN correspond to the true positive, false positive, and false negatives for each class labels, respectively. Weighted average precision (wAP) score is calculated by taking the weighted mean of the precision scores for each class, where the weights are the number of true instances for each class label.
\section{Experiments}\label{sec:exp}
\subsection{Experimental Setup and Datasets}

We pretrain our contrastive network on the unbalanced Audioset 
dataset and transfer it to sound-event classification for a subset of the Audioset and DESED \cite{Turpault2019} dataset. 
Audioset \cite{45857} consists of over 2 million 10-sec audio clips from YouTube videos. The clips are weakly labeled for over 500 classes. We don't use labels when pretraining the self-supervised network. This dataset has been used for pretraining SSL models in the literature \cite{fonseca2021unsupervised, niizumi2021byol, saeed2021contrastive}.

We perform two types of transfers for the downstream task: (1) Freezing the embedding and training a linear classifier on top, and (2) finetuning the entire network.

The mel-spectrograms calculated in the augmentation module use $25$ ms window size, $10$ ms hop-size, and $64$ frequency bins for $96$ frames. Following COLA, the encoder is EfficientNet-B0 \cite{tan2019efficientnet} and the projection head is a fully-connected layer with dimension $512$ and $tanh$ activation. We use ADAM \cite{kingma2014adam} with learning rate $10^{-3}$ and batch size $128$ for optimization.

\subsection{Results: DESED }
The DESED dataset \cite{Turpault2019} comprises 10 second audio clips recorded in domestic environment or synthesized to simulate a domestic environment, and focuses on 10 class of sound events. There are both real and synthetic recordings in the dataset. We used the strongly labeled real validation set for training the linear classifier or finetuning the entire network, and the evaluation set for testing.

Table \ref{tab:desed} shows the F1 scores on the DESED dataset. We present the results for both the cases where we freeze the encoder and train a linear classifier on top of the network, and when we finetune the entire network. We use COLA \cite{saeed2021contrastive} as the baseline. It's a special case of our framework in which only the noise block is used in the augmentation module. We consider different augmentation strategies and compare the corresponding F1 scores against the baseline and the supervised model. The results show that COLA with finetuning, having no mechanism to enforce feature invariance, was only slightly better than the supervised model. On the other hand, our finetuned model with time stretching and time/freq masking augmentation surpassed the full supervision by almost 4\% in F1 score. This highlights the importance of our augmentation module in this framework. Note that the low F1 scores of the SSL models are due to the distribution mismatch between the SSL data (Audioset) and the classification data (DESED).

\begin{table}[]
        \centering
        \caption{Average F-1 score for DESED dataset for different augmentation strategies.
        \label{tab:desed}}
        \begingroup
            \renewcommand{\arraystretch}{1.2} %
        \resizebox{\columnwidth}{!}{

        \begin{tabular}{ l c c }
        \toprule
                Augmentation type & Freeze Encoder & Finetune  \\
                \midrule
                Noise only (baseline) &0.434 &  0.629\\
                Time stretching & 0.454 & 0.640\\
                RIR filtering & 0.453 & 0.616 \\
                Time/freq masking & 0.492 & 0.624\\
                Time stretching + time/freq masking & \textbf{0.511} & \textbf{0.644 }\\
                Supervised & \multicolumn{2}{c}{0.62} \\
        \bottomrule
        \end{tabular}}
        \endgroup
\end{table}

\begin{figure} 
        \centering
        \includegraphics[scale=0.5]{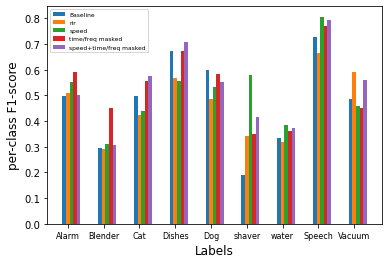}
        \caption{Per-class F1-score for DESED dataset.}
        \label{fig:f1-perclass-desed}
\end{figure}

Figure \ref{fig:f1-perclass-desed} presents the per-class F1-score for the considered augmentation combinations. Evidently, for almost all classes, there were one or more transformation combinations producing higher F-1 scores than that of the baseline. Furthermore, F-1 scores yielded by our best combination, time stretching + time/freq masking, were among the top scores for a majority of classes. This is consistent with Table~\ref{tab:desed} which shows that this strategy has the highest score on average for all classes.

\begin{figure}
\begin{tabular}{c c}
    \centering
    \includegraphics[scale=0.17]{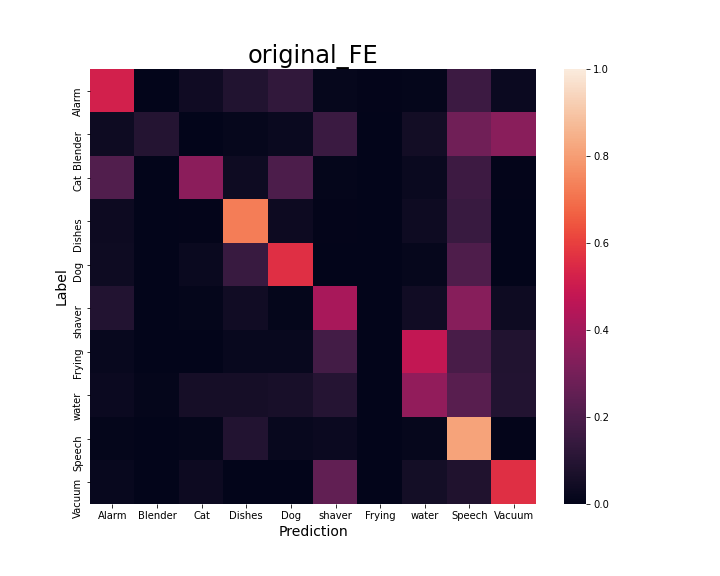}&
    \includegraphics[scale=0.17]{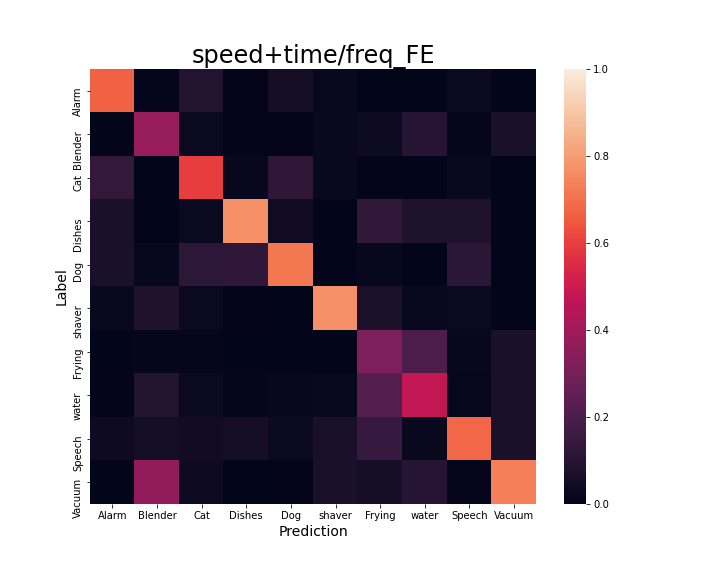}\\
    \includegraphics[scale=0.17]{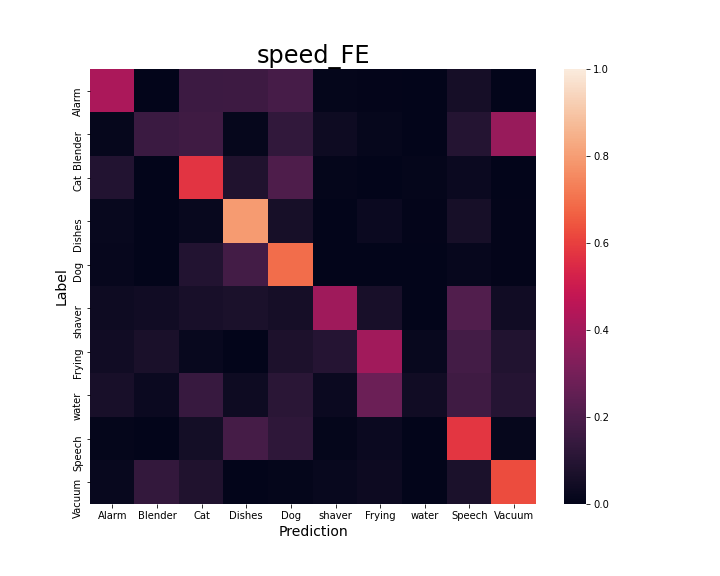}&
    \includegraphics[scale=0.17]{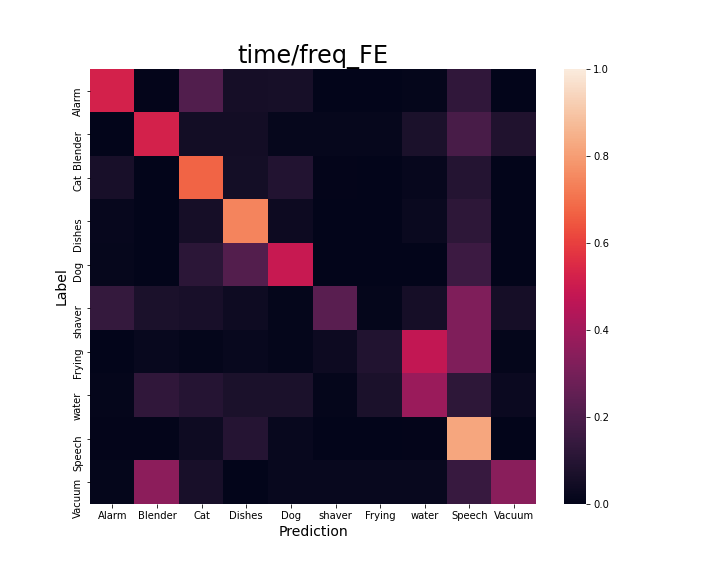}
\end{tabular}
\caption{Normalized confusion matrices for DESED dataset. \textit{Top left}: baseline. \textit{Bottom left}: time stretching. \textit{Bottom right}: time/freq masking. \textit{Top right}: time stretching + time/freq masking. }
\label{fig:confusion}
\end{figure}

Finally, in Figure \ref{fig:confusion}, we show the normalized confusion matrices for four augmentation types for the frozen encoder case. The normalization is applied over the total number of true labels for each class. That the non-diagonal elements of the time stretching + time/freq masking confusion matrix generally have lower values than those in other confusion matrices indicates that this strategy yields the best classification performance. This matches the insights from previous results.

\subsection{Results: Audioset}
\begin{table}[]
        \centering
        \caption{Weighted average precision score for Audioset dataset for different augmentation strategies. \label{tab:Audioset}}
        \begingroup
            \renewcommand{\arraystretch}{1.2} %
            \resizebox{\columnwidth}{!}{
            \begin{tabular}{ l c c  }
                \toprule
                    Augmentation type & Freeze Encoder & Finetune \\
                \midrule
                     Noise only (baseline) & 0.361 &  0.375 \\
                     Time stretching & 0.391 & 0.393 \\
                     RIR filtering & 0.384 & 0.385 \\
                     Time/freq masking & 0.378 & 0.395 \\
                     Time streching + Time/freq masking & \textbf{0.393} & \textbf{0.399} \\
                     Supervised & \multicolumn{2}{c}{0.371} \\
                \bottomrule
            \end{tabular}}
        \endgroup
\end{table}

We also assessed the representations learned by SSL pretraining using a subset of the Audioset. This subset includes 2000 weakly labeled sound clip for 10 most populated classes available. As the test set is imbalanced, we report, in Table~\ref{tab:Audioset}, the weighted average precision (wAP) scores for different augmentation types and compare them to those for the baseline (the COLA framework \cite{saeed2021contrastive}) and the supervised network. The results echo the phenomenon in Table~\ref{tab:desed} that COLA is inferior to the supervised network without finetuning and only slightly better than it with finetuning. Significantly, all of our augmentation strategies in our framework are superior to full supervision with and without finetuning. This indicates that, compared to naive contrastive SSL methods such as COLA, our augmented contrastive SSL framework learns more robust representations benifiting downstream tasks.

\section{Conclusion}
Improving generalization for audio classification requires solving the labeled data scarcity issue. In this work, we propose an augmented contrastive self-supervised learning method that, using unlabeled data, learns invariant representations benefiting downstream supervised tasks. Using the Audioset and DESED datasets, we show that our framework significantly outperforms state-of-the-art SSL and supervised learning methods on audio classification problems. Understanding how contrastive SSL combined with data augmentation produce features invariant to nuisance factors is still an open problem. We preserve it for future work.

\bibliography{ref}
\bibliographystyle{IEEEbib.bst}

\end{document}